\documentstyle[preprint,aps]{revtex}
\begin{document}
\draft
\title{ Electronic structure of FeSi }
\author{V.~R.~Galakhov, E.~Z.~Kurmaev, V.~M.~Cherkashenko,
Yu.~M.~Yarmoshenko, S.~N.~Shamin }
\address{ Institute of Metal Physics, Russian Academy of Sciences --
Ural Division, 620219 Yekaterinburg GSP-170, Russia }
\author{ A.~V.~Postnikov\cite{*}, St.~Uhlenbrock, M.~Neumann }
\address{ Universit\"at Osnabr\"uck -- Fachbereich Physik, D-49069
Osnabr\"uck, Germany }
\author{ Z.~W.~Lu, Barry M.~Klein, and Zhu-Pei Shi }
\address{ Department of Physics, University of California,
Davis, CA~95616-8677, USA } \date{\today}
\maketitle
\begin{abstract}
The full set of high-energy spectroscopy measurements including
X-ray photoelectron valence band spectra and soft X-ray emission
valence band spectra of both components of FeSi
(Fe $K_{\beta_5}$, Fe $L_{\alpha}$, Si $K_{\beta_{1,3}}$
and Si $L_{2,3}$) are performed and compared with
the results of {\it ab-initio} band structure calculations using the
linearized muffin-tin orbital method and linearized augmented plane
wave method.
\end{abstract}
\section{Introduction}
\label{sec:intro}

FeSi is a non-magnetic \cite{wyi63}, narrow-gap semiconductor
\cite{jwwwa67,ktidv68} at low temperatures.  Its magnetic
susceptibility $\chi(T)$ increases with temperature and passes through
a maximum at $T \sim 500$ K \cite{jwwwa67}.
FeSi becomes metallic above 300 K\cite{ktidv68,wwh65}.  The
substitution of Co for Fe (about 10\% Co) yields a magnet with a
helical spin order \cite{bvr83}.  Local density
functional\cite{hohe64} band structure calculations
\cite{mh93,fkd94,gpb94} give the correct value of the semiconducting
gap (about 0.1 eV), but can not explain the large magnitude of
$\chi(T)$. According to infrared and optical measurements
\cite{sfzmt93}, the gap of 50 meV is gradually filled with increasing
temperature, with new spectral weight which can not be explained
within the conventional band structure picture.  In connection with a
temperature-induced local moment, a model based on unified
spin-fluctuation theory was proposed in Ref.~\onlinecite{tm79} which explains
$\chi(T)$ using a model density of states (DOS).

In spite of the large number of publications devoted to measurements
of X-ray photoelectron spectra (XPS) and ultraviolet photoelectron
spectra (UPS) of FeSi\cite{slfsk89,kmmos92,cymty94,oal87,ksnik82,ssmfi94},
most measurements were performed using polycrystalline samples which do not
allow precise measurements on clean surfaces free of contamination.
In this paper, we present a full set of precise spectral measurements,
data, including XPS of the valence band and X-ray emission valence spectra
for both components of FeSi, which were obtained on the same single
crystal, and providing experimental information about the distribution of
total and partial DOS in the valence band.

The already published information on the calculated
electronic structure of FeSi presented in Ref.\cite{mh93,fkd94,gpb94}
reveals only total DOS and Fe $3d$, Si $3s$ partial DOS distributions
in Ref. \onlinecite{fkd94}, and Fe $3d$, Si $3p$
DOS for the CsCl-type structure (which is rather
different from that known for the bulk FeSi\cite{wyi63}) in
Ref. \onlinecite{gpb94}.  Because of this we performed a new set of
band structure calculations of FeSi by two independent methods --
linearized muffin-tin orbitals (LMTO) and linearized augmented plane
wave (LAPW) -- which give more detailed information about the total
and the Fe $3d$, Fe $4p$, Si $3s$, Si $3d$ and Si $3p$ partial
DOS distributions.

\section{Experimental}
\label{sec:exp}

The Fe $L_{\alpha}$ ($2p$--$3d4s$ transition) X-ray emission spectrum
was measured on the RSM-500 type X-ray vacuum spectrometer with a
diffraction grating ($N=600$~lines/mm and $R=6$~m) and electron
excitation. The spectra were recorded in the second order of
reflection by a secondary electron multiplier with a CsI
photocathode. The energy resolution was about 0.35--0.40 eV.  The
X-ray tube was operated at $V=4.6$~keV, $I=0.4$~mA.

The Si $K_{\beta_{1,3}}$ ($1s$--$3p$ transition) X-ray emission
spectrum was measured using a fluorescent Johan-type vacuum
spectrometer with a position-sensitive detector \cite{dckgo84}. The Pd
$L$-X-ray radiation from a special sealed X-ray tube was used for the
excitation of the fluorescent Si$K_{\beta_{1,3}}$ spectra. A quartz
$(10\bar{1}0)$ single crystal curved to $R=1400$~mm served as a
crystal-analyzer. The spectra were measured with an energy resolution
of approximately 0.2--0.3~eV.  The X-ray tube was operated at $V=25$~keV,
$I=50$~mA.

The Si $L_{2,3}$ ($2p$--$3s3d$ transition) X-ray emission spectra of
FeSi were taken from Ref. \onlinecite{kvswk92}, and the Fe
$K_{\beta_5}$ ($1s$--$4p$ transition) X-ray emission spectrum was
reproduced from Ref.~\onlinecite{kolo70}.

The XPS valence band spectra of FeSi were measured using a
Perkin-Elmer ESCA spectrometer (PHI 5600 ci, monochromatized Al
$K_{\alpha}$ radiation). The FeSi single crystal was cleaved in high
vacuum prior to the XPS measurements.  The XPS spectra were calibrated
based on the Au $4f$-spectra of Au metal ($E_b$=84.0~eV).

X-ray emission spectra have been brought to the scale of binding
energies with respect to the Fermi level using the binding energies of
the relevant initial (core level) states of the X-ray transitions as
measured by the XPS technique.  Corresponding binding energies are
$E_b({\rm Fe~}2p)=706.7$~eV, $E_b({\rm Si~}2p)=99.3$~ eV. The values
of $E({\rm Fe~}K_{\alpha_1})=6403.86$~ eV and $E({\rm
Si~}K_{\alpha_1})=1740.1$~eV were taken for comparison of Fe
$L_{\alpha}$ and Fe $K_{\beta_5}$, Si $L_{2,3}$ and Si
$K_{\beta_{1,3}}$ X-ray emission spectra of FeSi.

The measured XPS and X-ray emission spectra are shown in
Fig.~\ref{spec}.

\section{Details of calculation}
\label{sec:calc}

Electronic structure calculations have been performed for the cubic FeSi
structure (8 atoms/cell, space group $P2_{1}3$) as determined in
Ref.~\onlinecite{wyi63} and discussed in detail in
Ref.~\onlinecite{mh93}.  We have used the cubic lattice constant of
$a=4.493$ \AA, with Fe atoms occupying the $(0.1358,0.1358,0.1358)a$ and
equivalent positions of the $B20$ structure, while Si atoms occupy the
$(0.844,0.844,0.844)a$ and equivalent positions.

In the calculations using the tight-binding LMTO method \cite{tblmto},
we used space-filling atomic spheres of equal size on Fe and Si sites
($R=1.394$ \AA) and no empty spheres were introduced.  The
exchange-correlation potential as proposed by von Barth and
Hedin \cite{vbh} was used. DOS calculated by the tetrahedron
method over 470 {\bf k}-points in the irreducible part of the
Brillouin zone are shown in Fig.~\ref{spec}, and compared with the
experimental spectra.  Our calculated electron bands are very close to
those obtained by Fu {\it et al.}\cite{fkd94} using the augmented
spherical wave method, and our calculated DOS agree
well with those of Ref.~\onlinecite{fkd94,cfu}.  We found a direct gap of
0.14~eV at the $X$ point of the Brillouin zone and an indirect gap of
0.05--0.08~eV, in agreement with the
resistivity measurement data reported in Ref.~\onlinecite{fkd94}.

Some small deviations from the bands calculated in
Ref.~\onlinecite{mh93} using the LAPW method without any shape
approximation imposed on the potential seem to have quite negligible
effect on the densities of states, in the present context of making a
comparison with X-ray spectra using the LMTO method.
We also carried out an independent LAPW\cite{ande75,sing94}
calculation, in which we used the local density approximation
\cite{hohe64} form for exchange and correlation given by Ceperley and
Alder \cite{cepe80}, as parameterized by Perdew and
Zunger\cite{perd81}.  In the expansion of the charge density and
potential inside the muffin-tin spheres ($R_{\rm Fe}=R_{\rm Si}=1.111$
\AA), lattice harmonics up to angular momentum of $l=8$ have been
used, while the interstitial region was described by a plane wave
expansion.  Core orbitals were treated self-consistently, retaining,
however, only the spherically-symmetric part of their density. Scalar
relativistic effects (neglecting spin-orbit coupling) were included
for the valence states, and full relativistic effects were included
for the core states.  A large basis set consisting of both real-space
orbitals (inside the muffin-tin regions) and plane waves was used. The
total number of basis functions used was $\sim$ 570 per unit cell of 8
atoms.  The Brillouin zone summations were performed using 24 {\bf
k}-vectors in the irreducible section of the Brillouin zone in the
self-consistency loop.  The densities of states calculated using
the LAPW method, with the tetrahedron integration
over 176 {\bf k}-points in the irreducible
part of the Brillouin zone, are shown in
Figs.~\ref{dospart} and \ref{dostot}.
Note that the total DOS calculated in both LMTO
and LAPW methods per unit cell (i.e. 4 formula units) are in good
absolute agreement, whereas the partial DOS are somehow smaller in the
LAPW calculation (Fig.~\ref{dospart} because they are attributed
to non-overlapping muffin-tin spheres rather than space-filling
atomic spheres as in LMTO.
The expanded part of the total
densities of states plot in the vicinity of the gap is shown at the
inset in Fig.~\ref{dostot}. Two sharp peaks formed mostly by Fe $d$
states are separated by the indirect gap of about 0.065 eV.

There seems to be generally good agreement between the several LDA-based
band structure calculation schemes in reproducing the bands, total
densities of states, and bandwidths in FeSi. Below, we concentrate on
discussing particular features of partial densities of states of the
constituents as revealed in the X-ray emission spectra and the present
calculations.

\section{Experimental results and discussion}
\label{sec:disc}

The X-ray photoelectron valence band spectrum
and X-ray emission valence band spectra
of Fe (Fe $L_{\alpha}$, Fe $K_{\beta_5}$) and Si (Si $L_{2,3}$, Si
$K_{\beta_{1,3}}$) are presented in Fig.~\ref{spec}. As is known,
the XPS valence band spectrum gives an information about the total
DOS distribution (accurate up to the weight function depending
on the atomic photoemission cross-sections, see Ref.~\onlinecite{yl85}),
whereas the Fe $L_{\alpha}$, Fe $K_{\beta_5}$, Si $L_{2,3}$,
Si $K_{\beta_{1,3}}$ X-ray emission spectra (in accordance to dipole
selection rules) -- give information about partial Fe $3d4s$, Fe $4p$,
Si $3s3d$ and Si $3p$ densities of states, respectively.
These spectra are compared in Fig.~\ref{spec} with the results
of the LMTO band structure calculations by aligning the calculated
and experimental positions of the Fermi level. We note that
the experimental XPS valence band spectrum reproduces the total DOS
distribution of FeSi quite well.  Especially, it is very important
that in our experiments we have found the splitting of the main peak
of XPS spectrum in the range 0--2 eV which corresponds to the
Fe $3d$ band splitting.  This distinct splitting was not found before
in XPS\cite{slfsk89} and UPS\cite{kmmos92,cymty94,ksnik82,ssmfi94}
measurements on FeSi, and its absence was attributed in
Ref.~\onlinecite{ssmfi94} to a hole lifetime broadening which increases
with binding energies. The similar splitting of the XPS valence
band spectrum of FeSi was found in Ref.~\onlinecite{oal87} only
in the measurements done at low temperatures ($T$=120 K).
In our opinion, this discrepancy between fine structure in the present
and previous XPS (UPS) valence band spectra is due to the high quality
of the FeSi single crystal which was used for the present measurements.
The low-energy Si $3p$ and Si $3s$ subbands are reflected
in the XPS valence band spectrum as low-intensity
(due to low values of respective photoinization
cross-sections \cite{yl85}) features located at binding energies of
approximately 4.5 and 9 eV, respectively.

The experimental Fe $L_{\alpha}$ X-ray emission spectrum reproduces
fairly well the position of the center of gravity of calculated Fe
$3d$ DOS distribution in the valence band of FeSi, but
not the splitting of the Fe $3d$ band. This is due to the large total
distortion of the Fe $L_{\alpha}$ X-ray emission spectrum which
includes the instrumental distortion (about 0.4 eV) and the width of
the inner (core) Fe $2p$ level (about 0.8-1.0 eV) which is determined
by the lifetime of the core-level vacancy under an X-ray transition.

The experimental Fe $K_{\beta_5}$ spectrum shows two maxima located at
binding energies of approximately 3.5 and 9.0 eV that is in accordance
with the theoretical distribution of the Fe $4p$ partial DOS.
It is also seen from both the theoretical and experimental
spectra that Fe $4p$ states are hybridized mostly with Si $3p$ and Si
$3d$ states.

The experimental Si $L_{2,3}$ X-ray emission spectra corresponds to
the $2p$--$3s3d$ transition. Usually it is considered that the Si $3d$
states do not take part in the chemical bonding in $3d$ transition
metal silicides. As was emphasized in Ref.~\onlinecite{kvswk92}, the
high-energy subband of Si$L_{2,3}$ X-ray emission spectra of
transition metal silicides FeSi, MnSi and NiSi, as well as that of
disilicides FeSi$_2$, MnSi$_2$ and NiSi$_2$, cannot be explained
without an assumption that Si $3d$ states contribute to the chemical
bonding. Subsequently, the same conclusion was drawn for Pt silicides
(Pt$_2$Si, PtSi) in Ref.~\onlinecite{yhkin94} based on an analysis of
the Si $L_{2,3}$ X-ray emission spectra of these compounds. This
conclusion is again confirmed now for FeSi by two independent sets of
band structure calculations (see Figs.~\ref{spec} and \ref{dospart}).
One can see from Fig.~\ref{spec} that the Si $3s$ DOS is
concentrated in the range of binding energies about 6-13 eV where the
most intensive low-energy subband of the Si $L_{2,3}$ X-ray emission
spectrum is located. The center of gravity of the Si $3d$ states
distribution corresponds to a binding energy about 2 eV where the
high-energy maximum of the Si $L_{2,3}$ spectrum is situated.  The Si
$3d$ states are strongly hybridized with Fe3$d$ states, and the same
splitting of the Si $3d$ band (about 2 eV) as that in the Fe $3d$
subband is observed in the band structure calculations.

The energy position and fine structure of the Si $K_{\beta_{1,3}}$
X-ray emission spectrum is in a good accord with the Si3 $p$ partial
DOS distribution (see Fig.~\ref{spec}).

\section{Conclusion}
\label{sec:conclu}

The results of high-energy spectroscopy measurements of FeSi single
crystals, including the XPS valence band spectrum and the X-ray emission
valence band spectra of both constituents, are presented. They are
compared with two independent {\it ab initio} band structure
calculations of FeSi, performed using the LMTO and LAPW methods, and
good agreement between the experimental and theoretical spectra is
found.

\acknowledgements

The authors are grateful to Castor Fu for sending his data on the
calculated density of states of FeSi.  AVP, StU and MN appreciate the
financial support of the Deutsche Forschungsgemeinschaft (SFB~225,
Graduate College).  This work was supported by Russian Foundation for
Fundamental Research (grant No.94-03-08040) and NATO International
Scientific Exchange Program (Project HTECH LG 940861). ZWL, BMK, and
ZPS thank the support by the University Research Funds of the
University of California at Davis.

\newpage
\begin{figure}
\caption{ Measured XPS spectrum (dots; top panel) and X-ray emission
spectra (dots and dashed line for Fe $K_{\beta_5}$; lower panels) of
FeSi as compared with total (per unit cell) and partial
(per atomic sphere) densities of states calculated
using the LMTO method.  }
\label{spec}
\end{figure}

\begin{figure}
\caption{ Partial densities of states within muffin-tin spheres
at Si and Fe sites of FeSi
calculated using the LAPW method. Note that the scale for the Fe $3d$
DOS is different from the rest and that the Si $3d$ DOS has been multiplied
by a factor of ten (10).}
\label{dospart}
\end{figure}

\begin{figure}
\caption{ Total density of states (DOS) of FeSi calculated using the
LAPW method.  The insert shows an expanded view of the
DOS near the Fermi energy.
}
\label{dostot}
\end{figure}

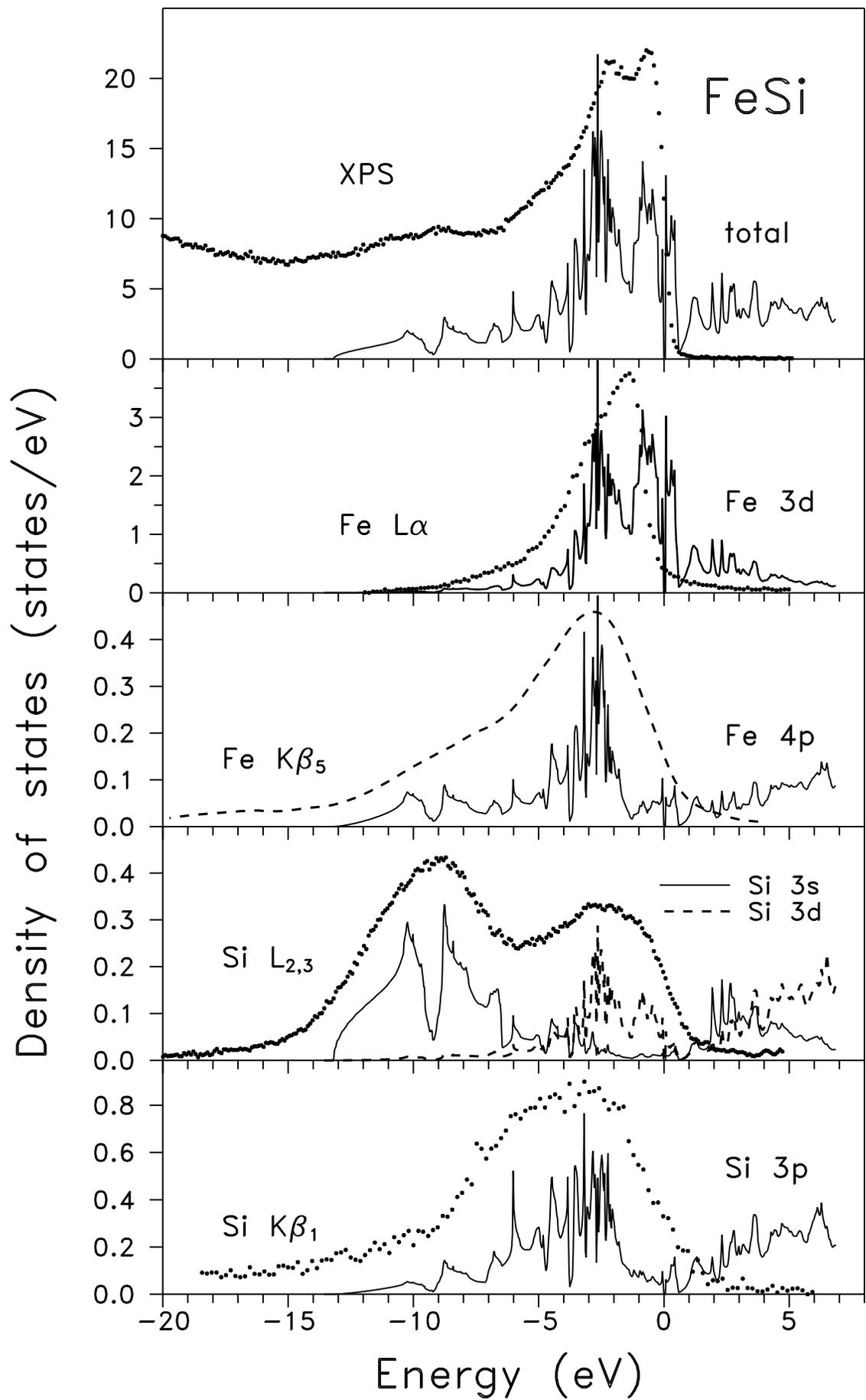
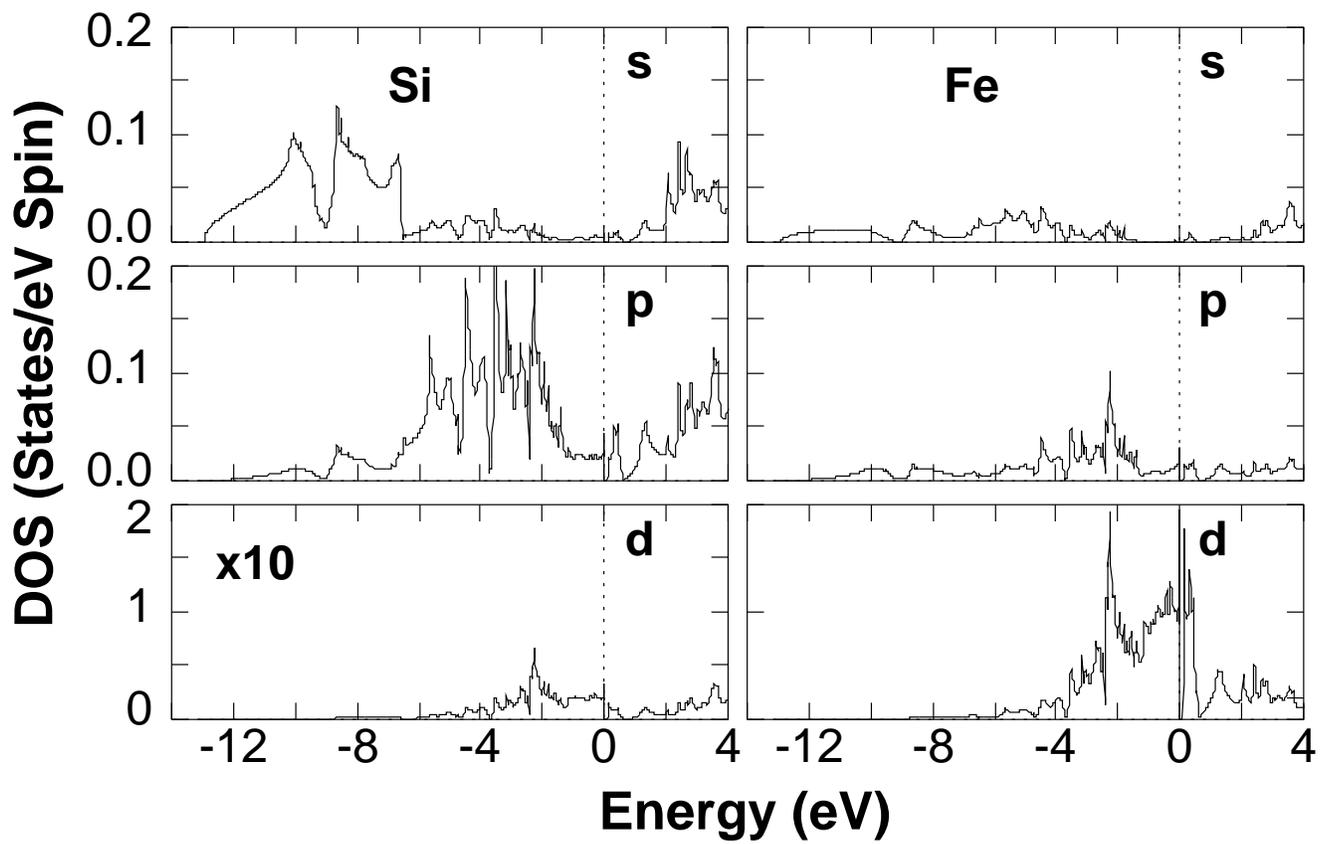
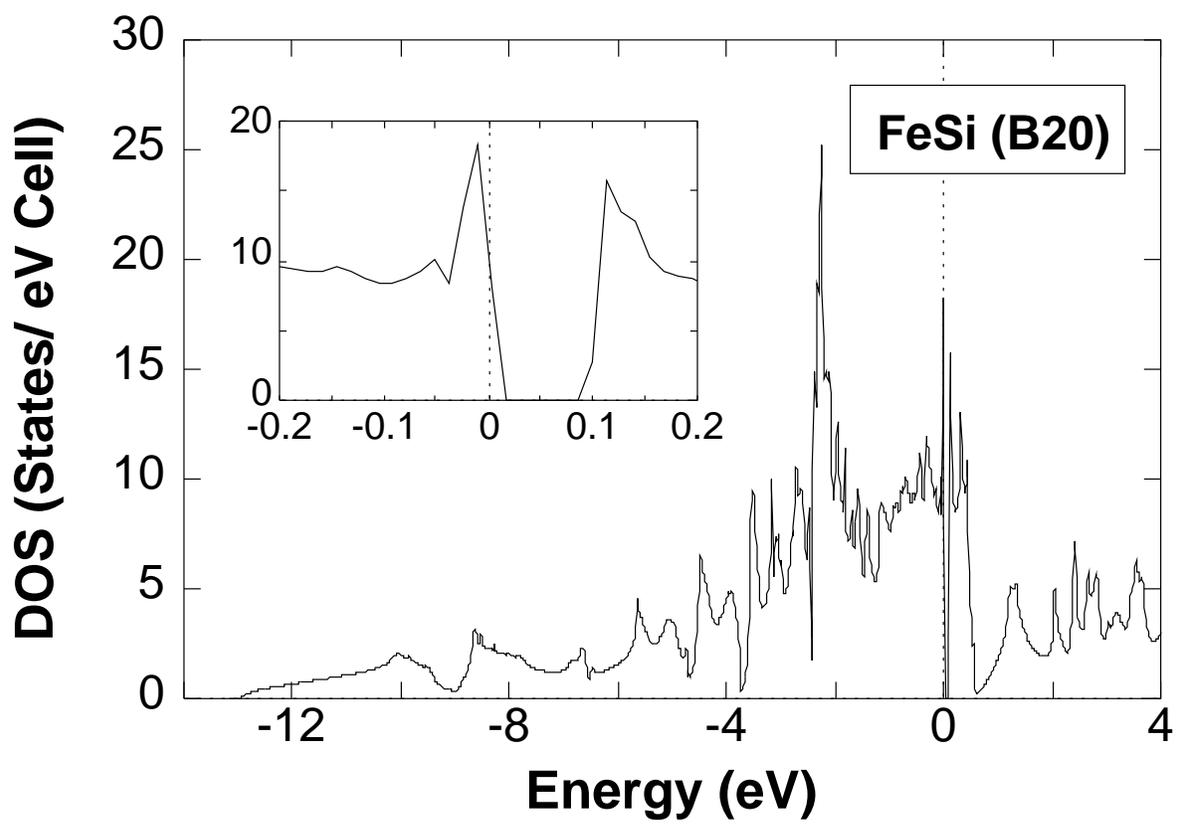
\end{document}